\documentclass[twoside,a4paper,11pt]{sca}
\usepackage{graphicx}
\begin{document}
\pagenumbering{arabic}
\pagestyle{myheadings}
\thispagestyle{empty}
{\flushright\includegraphics[width=\textwidth,bb=90 650 520 700]{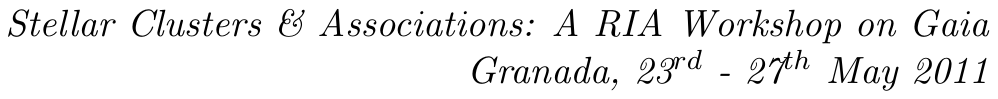}}
\vspace*{0.2cm}
\begin{flushleft}
{\bf {\LARGE
%
Birth, Evolution and Death of Stellar Clusters
%
}\\
\vspace*{1cm}
%
Richard de Grijs
%
}\\
\vspace*{0.5cm}
%
Kavli Institute for Astronomy and Astrophysics, Peking University, Yi
He Yuan Lu 5, Hai Dian District, Beijing 100871, China
%
\end{flushleft}
%
\markboth{
Star cluster lifecycles
}{ 
%
Richard de Grijs
%
}
\thispagestyle{empty}
\vspace*{0.4cm}
\begin{minipage}[l]{0.09\textwidth}
\ 
\end{minipage}
\begin{minipage}[r]{0.9\textwidth}
\vspace{1cm}
\section*{Abstract}{\small
%
Using our recently improved understanding of star cluster physics, we
are now within reach of answering a number of fundamental questions in
contemporary astrophysics. Star cluster physics has immediate bearing
on questions ranging from the physical basis of the stellar initial
mass function -- Do any O-type stars form in isolation? What is the
relative importance of stochastic (random) star formation versus
competitive accretion? -- to the build-up of the most massive clusters
-- Does the cluster mass function differ in different types of
galaxies? How and why do the most massive star clusters form in small
dwarf galaxies and what does that imply for the build-up of larger
cluster samples? What are the main observables one could (or should)
use to try and distinguish among the various star- and
cluster-formation scenarios? Newly emerging theoretical insights,
novel high-quality observational data and the advent of the next
generation of observational facilities offer significant promise to
reach satisfactory and robust answers to the key outstanding questions
in this field.
%
\normalsize}
\end{minipage}
%
%
%
\section{Preamble: Tertulias}

When I was first approached to lead a main `Tertulia' during the {\it
  RIA workshop on Stellar Clusters and Associations}, I was rather
puzzled by the invitation, to say the least. Although this expression
may be commonplace to some, it is not outside of the Spanish-speaking
diaspora. Fortunately, our trusted friend
Wikipedia\footnote{http://en.wikipedia.org/wiki/Tertulia} offered
answers: tertulias were originally informal, social literary or
artistic get-togethers, particularly in Latin cultural contexts, often
(but not always) held in public places. In relation to this
conference, however, the organisers meant me to chair an open
discussion on outstanding questions in contemporary
star cluster-related astrophysics. In the following, I aim at setting
the scene for the discussion that followed, which I attempt to
summarise in general terms. The lively exchange of new (and some old)
ideas that ensued led, I believe, to a general broadening of
participants' perspectives -- in the true sense of the traditional
tertulia.

\section{Outline of key emerging questions}

Stars do not form in isolation, at least for stellar masses above
$\sim 0.5$ M$_\odot$. In fact, 70--90\% of stars may form in star
clusters or associations (cf. Lada \& Lada 2003). Star formation
results from the fragmentation of molecular clouds, which in turn
seems to preferentially lead to star cluster formation. Over time,
their member stars become part of the general field stellar
population. Star clusters are thus often stated as being among the
basic building blocks of galaxies.

Using our improved understanding of star cluster physics, we are now
within reach of answering a number of fundamental questions in
contemporary astrophysics, ranging from the formation and evolution of
galaxies to the details of the process of star formation itself. These
two issues are the backbone of research in modern astrophysics. They
lead to new questions (for a detailed discussion, see de Grijs 2010),
including:
\begin{itemize}
\item How is star cluster formation triggered and how does it proceed? 
\end{itemize}

What is the role of ambient or internal pressure? Which other internal
and/or external factors affect star cluster formation and longevity?
Can we set constraints on the minimum requisite star-formation
efficiency? Is star cluster formation a distinct mode of star
formation or simply part of a broad spectrum of star-formation modes?

\begin{itemize}
\item Do star clusters really represent the basic building blocks of
  galaxies?
\end{itemize}

How does the star cluster mass distribution relate to the turbulent
properties of the interstellar medium (ISM)? Does initial substructure
play a role? Is there a physically important difference between star
formation in a `fractal' ISM versus in clustered mode? Are mergers of
smaller components a viable way to form more massive clusters? If so,
what are the constraints implied for the relevant parameter space
(e.g., velocity dispersions, half-mass radii, etc.?)

\begin{itemize}
\item How is the stellar mass distribution established?
\end{itemize}

Although the stellar initial mass function (IMF) seems fairly
universal, we still do not know what establishes its well-known
(multiple) power-law shape. What are the roles and the formation
scenario(s) of massive versus low-mass stars and the importance of
feedback? How are stellar and cluster IMFs related and how are they
affected by the underlying star-formation history?

\begin{itemize}
\item How do environmental conditions affect further cluster formation? 
\item How do quiescent galaxies form extremely massive clusters?
\end{itemize}

What is the role of galaxy dynamics? Does it depend on a cluster's
position in a given galaxy or on galaxy type?  How are the results
affected by variations in the star-formation efficiency, or are we
simply witnessing stochastic cluster formation? What triggers the
formation of the most massive clusters?

These (and other) questions form the basis of much research in
numerous fields in contemporary astrophysics: after all, star
formation is one of the pivotal physical processes underlying most of
astrophysics. Significant progress has been made in past decades, yet
we still have a long way to go before all of the key questions will
have been answered satisfactorily and sufficiently robustly to stand
the test of time. 

New advances in both theoretical and observational approaches may be
required. Are the main theoretical (modelling) challenges preventing
the next step change in our understanding of star and star cluster
formation processes related to limitations in hardware or techniques?
Are currently available instruments sufficient to reach these lofty
goals, or will new facilities -- including {\sl Gaia}, the Atacama
Large Millimetre Array (ALMA), the {\sl James Webb Space Telescope},
the Large Synoptic Survey Telescope, or any flavour of extremely large
telescope -- be essential? Do we need higher spatial resolution than
currently routinely attainable or perhaps larger fields of view, or a
combination of both -- and is this feasible?

\section{Are star clusters indeed basic galactic building blocks?}

The question as to what constitutes a characteristic scale in star
formation comes down to our understanding of the way in which stars
appear to form in a self-similar, hierarchical (`fractal')
fashion. Observationally, we find a wide range in the gas distribution
and star-formation properties in regions of active star formation
(e.g., Bressert et al. 2010), where what we tend to call `clusters'
are the loci in which many stars form together in the basic,
underlying hierarchical scenario. {\sl Gaia} may well be instrumental
in refining our understanding of the scales on which star formation
occurs in the Milky Way.

Turbulent molecular clouds form clusters in the denser regions, where
filaments might cross. Star formation occurs along such filaments,
usually in small groups (`knots'), which is shown beautfully by recent
observations with {\sl Herschel}. This is sometimes referred to as the
distributed mode of star formation, but it is an integral part of the
power-law distribution of structures. Clusters, or `spatially and
temporally correlated star formation' follow a power-law distribution
of mass at a given time. One can start building galaxies from that
Ansatz.

In some places, large numbers of stars are form at roughly the same
time. These loci are easily identified as star clusters. At other
times, star formation proceeds in a more distributed mode, upon which
the individual knots merge and also form structures we identify as
star clusters. {\sl Gaia} will likely find a lot of evidence of
distributed star formation, where most stars never originated in
clusters. The European Space Agency's mission will, therefore, be key
in distinguishing between these models.

Superficially, it appears that two somewhat different views are
prevalent in the community, apparently advocating the extremes of our
ideas of how stars form and the structures resulting from that
process. Proponents of one viewpoint would argue that star formation
occurs in a spatially confined, clustered mode, while their
counterparts support the full hierarchical picture. However, as long
as proponents of the tenet that star clusters are the basic building
blocks from which galaxies form take their view down to the
lowest-mass, often embedded clusters (which includes the distributed
population of young stars), these apparent differences are merely
semantic.\footnote{At the meeting, this was effectively summarised by
  Pavel Kroupa, who commented -- in response to an argument proposed
  by Simon Goodwin -- that ``[w]hat I said is essentially what you
  said, just a bit more mathematical.''}

In essence, therefore, it is not important whether the star-formation
process forms a dense cluster, an embedded cluster or an
association. What matters is whether star formation is correlated: are
star-formation events quantised? Is there any length scale that is
special and unique? If so, clusters of a given size and/or mass may
well be the basic building blocks of galaxies. However, if star
formation is hierarchical, this implies that there is no unique
timeframe or spatial scale on which star-forming clumps are absolutely
correlated. In the latter case, star clusters and associations are not
unique; instead, they simply represent a continuous distribution. In
extragalactic environments, at least, on spatial scales in excess of 5
pc there is no specific length scale (e.g., Bastian et al. 2009,
2011).

However, there is a unique scale at bottom of the fractal distribution
in both the stellar and gas distributions, i.e., that of a dense core:
$\sim 0.1$ pc. This is the scale of individual stars, which we could
potentially resolve with ALMA or -- for nearby star-forming regions --
the Plateau de Bure interferometer. This is of a similar order as the
scale of subsonic turbulence. In addition, the IMF exhibits a peak at
stellar masses between $\sim 0.5$ and 1 M$_\odot$, which also implies
a characteristic {\it mass} scale.

From this discussion, the following hypothesis naturally results. The
fundamental galactic building blocks are units of star formation of a
few tenths of a parsec across, distributed according to a mass
spectrum that is a power law down to a few M$_\odot$. If one were to
distribute this kind of structure throughout a region of a galaxy
according to a model of what the ISM looks like -- essentially a
fractal distribution -- and then let it evolve dynamically, the result
would enable us to derive which fraction of stars is found in
distributed versus clustered mode. The null hypothesis of a scale-free
star-formation process can thus be either verified or discarded.

In a hierarchical scenario, this picture would be created by a
turbulence power spectrum that is self-similar. However, even in such
an ISM there is a definite scale, i.e., the scale on which the
turbulence is injected into the ISM -- and this differs among galaxies
because of variations in the perturbing mechanism. One cannot have
correlated structures on scales that are larger than the scales on
which the energy is injected. In Holmberg 2, a dwarf galaxy, Dib \&
Burkert (2005) found that the structures are correlated on a very
large scale of $\sim 6$ kpc. In the Galaxy, Dib et al. (2009) found
that the orientations of molecular clouds are spatially correlated on
scales of a few 100 pc. This may be related to some instability in the
outer Galaxy, because it is on the order of the scales of supernova
remnants which evolve in a low-pressure environment. The conclusion
is, therefore, that at least on large scales there must also be a
characteristic spatial scale.

From the preceding discussion, it seems that asking whether star
clusters are basic galactic building blocks is a matter of
semantics. More interesting, perhaps, is understanding why, when and
how star formation happens. Star clusters tell us about stellar
populations, the conditions in molecular clouds and the gas at the
time of their formation.

\section{How is the stellar mass distribution established?}

\subsection{Do any O-type stars form in isolation?}

Can O-type stars form in complete isolation? The answer to this
question is intimately linked to our understanding of where the IMF
comes from and how it fits in with the cluster mass--maximum stellar
mass relation (e.g., Weidner et al. 2010, and references therein). If
the (molecular cloud) core mass function is the origin of the stellar
mass function, then at the top end of the core mass function a very
massive core may occasionally be formed because of stochastic effects
(although note that observations are affected by small-number
statistics). In very rare circumstances, this core will not fragment
and collapses into one single massive object. For this to happen, the
core needs to be warm so that it does not fragment, although a small
cluster could form at the same place as well. In this case, the term
`isolated O stars' does not imply that no other stars can form
nearby. On the contrary, a small cluster, possibly containing an
O-star binary, might form, but the IMF is simply not fully sampled: it
may contain a single O star, a few G stars, and possibly one B star,
for instance.

If, on the other hand, the process of competitive accretion dominates,
large cores will fragment into many objects. In this case, the
starting point consists of cores with masses of around 1
M$_\odot$. These allow formation of a few objects per core, some of
which may then accrete additional mass. Proponents of this scenario
assume that most of the stellar mass function is set by the core mass
function, and most cores do not grow very massive. All cores start
small and if there is not much ambient gas, none can grow big. The
only way in which a massive core can develop is if there is a lot of
ambient gas present and a few objects can accrete a significant
fraction of that gas. Competitive accretion naturally leads to the
cluster mass--maximum stellar mass relationship. 

In reality, a combination of processes is likely responsible for the
final stellar mass distribution. For instance, in 30 Doradus (the
largest H{\sc ii} region in the Local Group), competitive effects must
have played a very significant role, but in associations this may be a
different side of the same coin. Perhaps we should instead ask the
question as to how frequently one or the other mode of star formation
dominates.

In the first scenario outlined above, the core mass function sets the
stellar mass function. In the latter case, however, one cannot really
predict the resulting mass function, yet the observed mass functions
are very similar. So, the simple question is whether we can quantify
any differences: Are there measurable differences between the mass
functions of cores and stars?

\subsection{What sets the initial mass function?}

The IMF and core mass function appear very similar in terms of their
morphology (except, perhaps, at the high-mass end), but this result
depends on the implicit assumption of one-to-one mapping from cores to
stars. However, we definitely observe cores (e.g. B59: Covey et
al. 2010) that are single, do not exhibit any substructure or
fragmentation, but contain 20 stars. Therefore, the scenario in which
we shift the core mass function to the stellar mass function (modulo
the star-formation efficiency, usually assumed to be $\sim 30$\%)
depends on some assumption as to how this efficiency acts on
individual cores. We have at least one example where the peak should
shift by more than a factor of 3. This problem is difficult to solve,
because once the stellar mass function can be measured, significant
stellar evolution is (and has already been) proceeding, while cores
represent the very early, almost unevolved stages of star formation.

In an alternative approach, one may explore differences in the mass
functions at the high-mass end, where any differences will be most
pronounced. Based on observations of a statistically large number of
clusters, it is very difficult to show that the observed, strict
relationship between the mass of the most massive star and that of its
parent cluster is valid under any circumstances. However, if one were
to observe the formation of a single massive star, that would
represent a significant deviation from that relationship. Therefore,
if a truly isolated O star were found (i.e., an exception to the
rule), there would be at least one case in which competitive accretion
has not worked, thus providing support for the stochastic sampling
theory.

In observational terms, let us compare Taurus, a region of active star
formation harbouring a mass of some $10^3$--$10^4$ M$_\odot$, and NGC
3603, a larger star-forming region. In essence, the random sampling
scenario implies that a collection of molecular material composed of
Taurus-like units equivalent to that of a large cluster would never
produce a massive star. However, proponents of the idea of competitive
accretion would argue that a large number of Taurus-like units {\it
  do} lead to the formation of massive stars. This apparent conflict
may disappear if it were a simple matter of scaling up one's
star-forming regions. If one considers Orion, for instance, the
morphology is always the same, exhibiting numerous filaments and
clumps -- just like the structure of Taurus. Scaling this kind of
hierarchical structure up leads to configurations resembling Orion
A. Alternatively, small sections of Orion A at high resolution look
exactly like Taurus or the Pipe Nebula.

An alternative approach to the conundrum of massive-star formation may
be found in considering the ratio between the number of massive to
low-mass stars in different regions. If we first consider Orion and
count the total number of low-mass stars as well as the number of
massive stars, and we find that the number of massive stars is
underrepresented in the entire Orion area, what do we deduce from
this? Similarly, if we explore this ratio in dwarf galaxies, i.e., we
count the number of massive stars in a dwarf galaxy with a low
star-formation rate and we find that there are too few massive stars
compared to what we expect from the shape of the `canonical' IMF, what
do we conclude from this, since it is all driven by purely stochastic
(random) sampling? The underlying idea of stochastic sampling is that
if we add many Taurus--Auriga-like structures to form an Orion-like
configuration, we should find an IMF that is more dominated by
low-mass stars than expected for a canonical IMF. We can thus either
exclude or verify random sampling as a viable star-formation scenario.

However, these scenarios may be too simplistic. First, if massive
clusters form from hierarchical mergers of subclusters before gas
expulsion, then the prestellar clumps in these subclusters do not know
{\it a priori} where they will end up. Why then is there a clear
correlation between the maximum stellar mass in a cluster and the
total cluster mass? Second, one may wonder whether there are there any
clumps that collapse in which the star formation has completely
stopped. Instead, star formation is likely ongoing while the clumps
collapse and during the early dynamical evolution of the resulting
systems. This early collapse occurs on essentially the free-fall
timescale, at least in Orion-like clusters (e.g., Allison et al. 2009,
1010; Yu et al. 2011). This is corroborated by arguments based on
structural parameters: at longer wavelengths (e.g., as seen with the
{\sl Spitzer Space Telescope}), star-forming regions are rather
filamentary. However, when we consider nearly `complete' clusters,
they are quite spherical but not extremely substructured: collisions
of these clumps seem, therefore, quite fast. Perhaps some of these
clumps first merge before forming a massive star (e.g., to release
angular momentum). One should keep in mind that the entire picture is
extremely dynamical; one certainly cannot assume that young
star-forming clumps evolve in isolation.

Finally, one should be cautious in linking the stellar mass
distribution directly to the star-formation process. Although we have
assumed in this discussion that the IMF is somehow the result of star
formation, it is perhaps better to state that at the end of the
star-formation process, a near-invariant IMF results. A similar IMF is
also observed at the start of the process of star formation. However,
one should keep in mind that the start of the fragmentation process
may be driven by different physics than that underlying that of star
formation itself (i.e., the reasons for the onset of the collapse). As
a consequence, depending on how one defines one scale, the resulting
clusters and cluster masses may differ, so that the IMF does not carry
much or any information about the star-formation process.

\subsection{Binarity}

What is the role of multiplicity in high-mass stars? There are
numerous examples of significant multiplicity fractions in young
(star-forming) environments, e.g. the Orion Nebula's Trapezium system
or the young Large Magellanic Cloud (LMC) cluster NGC 1818 (Hu et
al. 2010). Each of the massive Trapezium stars is at least a binary,
and there are many more examples. Even based on interferometric
submillimetre observations, some Class 0 objects have been found to be
multiple.

Let us assume that the power-law stellar mass distribution goes down
to very low masses. For stellar masses below $\sim 100$ M$_\odot$, the
population of these objects will not be dynamically processed; it will
simply disperse. Given that the IMF is quite invariant, it then
follows quite reasonably that the binary populations may also be
initially invariant (assuming universal binary properties). One can
now calculate the binary population expected for a galactic field or
an entire galaxy, if we understand star formation to the extent we
believe we do in the Milky Way, i.e., where the IMF and binary
populations are both universal and stars form in power-law
structures. One can show that the galactic-field binary population is
derived beautifully, because the more massive objects break up the
binaries, while the lower-mass objects do not. This leads to an
invariant IMF and binary population. One can then predict that dwarf
galaxies should have high and starburst galaxies low binary
fractions. This could potentially be checked observationally, at least
if we could resolve nearby galaxies into individual stars.

\section{What are the conditions for massive-cluster formation?}

On larger scales, how do globular clusters (GCs) form? The answer to
this question is linked to how environmental conditions affect further
cluster formation. Large numbers of massive proto-GCs formed in the
past, but why do we not see similar numbers at the present time in the
Milky Way? The massive young cluster Westerlund 1 may be a recent
example, although it is also possible that it may turn into an open
cluster or dissolve into the Galactic field because of its rather
extreme environmental conditions. In fact, we have to be careful in
making a clear distinction between open and globular clusters. We see
young GCs in massive galaxy mergers, and even in spiral disks. Instead
of advocating distinct formation mechanisms, we are most likely
observing the remnants of a much more varied original population --
which may have formed following a variety of mechanisms -- of which
many members have disappeared because of dynamical evolution. The main
physical driver underlying the mass build-up of young star clusters is
the star-formation-rate density: higher densities allow a cluster to
assemble more mass before star formation ceases.

At least some of the massive globular-like clusters in galaxies like
the Milky Way originated in smaller dwarf galaxy companions
(including, most likely, $\omega$ Centauri), which later merged with
the main galaxy dominating the local gravitational potential. This
leads to the question as to why we do not observe many of such massive
proto-GCs in dwarf galaxies like the Magellanic Clouds. The LMC
contains the most significant H{\sc ii} region in the Local Group, 30
Doradus, but it is small by comparison of the structures required to
form true GCs. Other nearby dwarf galaxies, including NGC 1569 and NGC
1705, do contain the type of object that may evolve into genuine GCs,
however.

How can such small dwarf galaxies host several globular or young
massive clusters that contain some 10\% of the total mass of these
galaxies? Molecular clouds can be larger in dwarf galaxies because of
the lower shear and tidal forces in these systems compared to large
spiral galaxies. In the Milky Way, molecular cloud masses are of order
$10^5$--$10^6$ M$_\odot$, while in the LMC they are approximately
$10^7$ M$_\odot$, for instance. Molecular clouds in galaxies with
lower shear can grow larger by agglomeration.

Two other, complementary effects (both related to metallicity) might
make clumps and/or giant molecular clouds more massive in galaxies
like the Large and Small Magellanic Clouds. In lower-metallicity
galaxies, the gas-to-dust coupling occurs at a different density
compared to that in higher-metallicity environments. The probability
density function is broader and extends to higher densities, because a
cloud has to accumulate more mass to shield itself (e.g., Glover \&
Clark 2011). Second, lower-metallicity feedback is less effective and
allows for more star formation, potentially also resulting in more
massive clusters and associations.

The natural question to ask in this context is whether the cloud (and
cluster) mass function in dwarf galaxies is different from that in
$L^\ast$ galaxies. Current thinking implies that complete universality
is no longer fully supported by the data, but one has to look
carefully at where the differences arise. There are no significant
differences between the Magellanic Clouds and the Milky Way. On the
other hand, despite the small-number statistics affecting these
arguments, it is curious that dwarf galaxies like NGC 1569 and NGC
1705 host very massive $> 10^5$ M$_\odot$) clusters, which are truly
outstanding in their respective cluster mass functions.

Ongoing research aimed at addressing the question of the origin of the
most massive young and globular clusters has not yet produced a
satisfactory answer. The most massive clusters in extragalactic
environments are found in merging systems. Observations and theory
support many mergers in the young Universe. However, mergers are not
necessarily the evolutionary end point of a galaxy: disks can reform
within a Gyr, potentially allowing the formation of new populations of
GCs at that stage.

Perhaps the most promising approach to addressing the GC formation
scenario resides in the use of concrete, observable indicators that
can help distinguish the various formation scenarios proposed. Massive
ellipticals contain the largest populations of GCs, the most massive
GCs and the highest metallicities. Dwarf galaxies have very low
metallicities, so that it appears unfeasible to merge large numbers of
dwarf galaxies to form ellipticals. To understand GC formation, much
better metallicity measurements are required. Similarly, better
determinations of ages and age spreads are advisable, although they
are much harder to obtain. Metallicities -- both integrated and of
individual stars in resolved, nearby clusters -- are the most basic,
clearest and most important indicator of how GCs may have formed.

Despite a significant body of recent work in this field, there is no
single observation of a subclustered massive proto-GC. Most of our
theoretical understanding of the details of massive cluster formation
hinge on scenarios involving scaled-up versions of smaller open
cluster-like configurations, where we do see clear substructure at
early stages. Nevertheless, tantalising objects that might indeed
provide clues as to the morphology of forming proto-GCs on small
scales include Sagittarius B2 (e.g., Bally 2010) -- observed as part
of the Bolocam Galactic Plane Survey -- and the embedded massive young
clusters in NGC 5253. Further Galactic Plane surveys (e.g., Lucas et
al. 2008; Minniti et al. 2010) will be instrumental in constraining
the early evolutionary properties of such objects.

Although at the present time we do not know of any proto-GCs in the
Milky Way, with the possible exception of Westerlund 1, our
observational sample is very incomplete: to date, only 14 massive
young stellar clusters have been discovered on the near side of the
Galaxy, all with masses around $10^4$ M$_\odot$. In addition, Davies
et al. (2011) recently discovered a 1 Gyr-old, $10^5$ M$_\odot$
cluster in the Galactic disk at a distance of only $\sim 2$--3
kpc. Indeed, very massive clusters do form occasionally in galactic
disks (e.g., Larsen et al. 2001).

%
\section*{Acknowledgments}   
%
It is my great pleasure to thank the organisers of this workshop for
the opportunity to shape this tertulia. In addition, I would like to
pay tribute to all active participants, without whose contributions we
would not have had such a lively exchange of ideas!

%

%
\end{document}